\documentclass{ws-procs975x65}            
\newcommand{\lambdabar}{{\mkern0.75mu\mathchar '26\mkern -9.75mu\lambda}}
\begin{document}
\title{Gravitational waves in a molecular medium:\\
dispersion, extra polarizations and\\
quantitative estimates}

\author{F. Moretti}

\address{Physics Department, “Sapienza” University of Rome,
	P.le Aldo Moro 5\\
 00185 Roma, Italy\\
E-mail: fabio.moretti@uniroma1.it
}

\author{G. Montani}

\address{ENEA, Fusion and Nuclear Safety Department, C. R. Frascati,
	Via E. Fermi 45\\
 00044 Frascati (Roma), Italy\\
 Physics Department, “Sapienza” University of Rome,
 P.le Aldo Moro 5\\
 00185 Roma, Italy
}

\begin{abstract}
We analyze the propagation of gravitational waves in a molecular matter medium: our
findings demonstrate that dispersion only is expected, together with the emergence of
three extra polarizations, able to induce longitudinal stresses. We also give quantitative
estimates of the predicted effects.
\end{abstract}


\bodymatter

\section{Introduction}

The Theory of General Relativity provides us a very accurate and elegant theoretical framework from which forecast on all natural phenomena dominated by gravity.
One of the most fascinating features of the theory is the prediction of the fact that the gravitational field is radiative: curvature can propagate through spacetime with the speed of light. With the first direct measurement of a gravitational wave, in 2015, we are witnessing the birth of a new era in the fields of astronomy, astrophysics and cosmology. The major part of theoretical studies have dealt with the asymptotic properties of gravitational waves, far from bounded sources \cite{a,b,c,d}. Many other authors have dealt with the problem of the propagation of gravitational waves in a matter medium: the case of a dissipative fluid is considered in \cite{hawking,madore,prasanna}, whilst a collisionless kinetic gas is studied in \cite{igna,polnarev,chesters,weinberg,flauger}; the interaction between gravitational waves and neutrinos is analyzed in \cite{lamo,labemo,belamo} whereas the case of spinning particles is investigated in \cite{milamo}.

Our aim is to analyze the propagation of gravitational wave in a matter medium described as a set of point-like masses, grouped into molecules: it has been shown\cite{e} that such a medium can be modelled, through the application of Kaufman molecular moments averaging method\cite{f} (whose application is denoted by $\left< \cdot \right >$), as
\begin{equation}
\left < T_{\mu\nu} \right > = T_{\mu\nu}^{(f)}+\dfrac{c^2}{2}\;\partial^\rho \partial^\sigma{Q}_{\mu\rho\nu\sigma },
\end{equation}
where $T_{\mu\nu}^{(f)}$ is the stress energy tensor that describes a set of free particles, \textit{i.e.} the centers of mass of the molecules, and $Q_{\mu\rho\nu\sigma}$ is the quadrupole polarization tensor, that gives account of the molecular structure, characterized by the same set of symmetries of Riemann tensor. 

We find, through a direct calculation performed via the geodesic deviation equation, that the quadrupole tensor is generated, on a static level, by the tensor of the second spatial derivatives of the Newtonian potential of the molecule. Moreover, if the molecule is perturbed with a vacuum gravitational wave we find that the latter is source of an induced contribution to the quadrupole moment of the molecule: the alteration in the structure of the molecules causes the fact that the medium reacts to the gravitational wave with a mean field that modifies the propagation of the wave inside the medium itself.

We show that only dispersion is expected, along with the appearance of three extra degrees of freedom able to induce longitudinal stresses on test particles. We also give some quantitative estimates of the predicted effects, by assuming the medium to be either similar to our own galaxy or to a more exotic object.
\section{Constitutive relation and modified wave equation}
The derivation of the constitutive relation is made under the hypothesis of slow motion of the point-like masses composing the molecule and small departure from Minkowski background metric. Moreover, the molecule is imagined as a sphere of constant mass density, characterized by a definite macroscopic radius $L$: the reduced wavelength $\lambdabar$ of the gravitational radiation interacting with the molecule must be greater than the molecular size. We perturb the molecule with a vacuum, plus polarized gravitational wave and we express the quadrupole tensor in terms of the amplitudes of the wave. We report the constitutive relation that we found:
\begin{equation}\label{matrela}
Q_{i0j0}=\epsilon_g \left(\dfrac{2}{c^2}\,\phi,_{ij}+\dfrac{1}{2}\,h_{ij,00} \right) \qquad , \qquad Q_{0ijk}=Q_{ijkl}=0.
\end{equation}
Here $\phi$ is the Newtonian potential of the molecule and $h_{\mu\nu}$ is the small departure from Minkowski background, the dynamical metric perturbation called gravitational wave. The constant $\epsilon_g$ is a real gravitational dielectric constant and can be calculated as
\begin{equation}
\epsilon_g=\dfrac{NL^5c^2}{4G},
\end{equation}
being $N$ the density of molecules inside the medium.
We write the macroscopic form of Einstein field equation
\begin{equation}
G_{\mu\nu}= \chi \left ( T_{\mu\nu}^{(f)}+\dfrac{c^2}{2}\;\partial^\rho \partial^\sigma{Q}_{\mu\rho\nu\sigma }\right) \qquad , \qquad \chi=\dfrac{8\pi G}{c^4}
\end{equation}
 at first order in $h_{\mu\nu}$, implementing the constitutive relation \eqref{matrela}: we find a set of fourth order partial differential equation for the wave inside the medium. We write the problem in terms of the trace reversed matrix $\bar{h}_{\mu\nu}$ on which we are allowed to activate Hilbert gauge fixing, given the diffeomorphism invariance of the theory:
 \begin{equation}
\bar{h}_{\mu\nu}=h_{\mu\nu}-\dfrac{1}{2}\eta_{\mu\nu}h \qquad , \qquad \partial^\mu\bar{h}_{\mu\nu}=0.
 \end{equation} 
 We find that the trace $\bar{h}$ solves
 \begin{equation}
 \Box \left ( \partial_0^2+m^2\right)\bar{h}=0,
 \end{equation}
 where $\Box$ is d'Alembert wave operator and $m^2=\dfrac{c^2}{4\pi G \epsilon_g}$: hence $\bar{h}$ is a superposition of an harmonic part, solution of d'Alembert equation, that can be canceled out with a further gauge transformation that preserves Hilbert gauge, plus a solution of the harmonic oscillator $\partial_0^2+m^2$ that does not propagate. This means that we can set $\bar{h}$ to zero: the remaining five degrees of freedom solve the fourth order PDE
\begin{equation}\label{eqmezzo}
\partial_0^4\,\bar{h}_{\mu\nu} + m^2 \Box\,\bar{h}_{\mu\nu}=0.
\end{equation}
We search for plane wave solutions of \eqref{eqmezzo}, choosing the $z$ axis to be coincident with the direction of propagation of the wave, $\bar{h}_{\mu\nu} \propto e^{i \left( \omega(k)t-kz\right)}$ and we calculate the dispersion relation:
\begin{equation}
\omega^2_{\pm}(k)=c^2 \bigg( -\dfrac{m^2}{2}  \pm \sqrt{\dfrac{m^4}{4}+m^2k^2} \bigg ).
\end{equation}
It can be shown that only the plus signed solution has a physical significance, because it goes back to be the standard vacuum dispersion relation $\omega(k)=ck$ when the material medium is removed, in the limit $\epsilon_g\to 0$, or $m^2 \to \infty$, in contrast to the minus signed, that shows a divergence in the same limit. It is also easy to show that $\omega_+(k)$ is real for any value of the wavenumber: dispersion only is expected. We calculate the group velocity
\begin{equation}
v_g(k)=\dfrac{m^2 kc}{2\sqrt{\left(\sqrt{m^2k^2+\dfrac{m^4}{4}}-\dfrac{m^2}{2}\right)\left(m^2k^2+\dfrac{m^4}{4}\right)}} \xrightarrow[k \ll m]{} c\left(1-\dfrac{3k^2}{2m^2} \right).
\end{equation}
We will show that the condition $\lambdabar \gg L$ always implies $k \ll m$.
\section{Polarizations}
We exploit the gauge freedom, fixing
\begin{equation}
\bar{h}=\bar{h}_{11}+\bar{h}_{22}+\left( 1-\dfrac{c^2k^2}{\omega^2}\right)\bar{h}_{33}=0,
\end{equation}
via the following equations:
\begin{equation}
\begin{split}
\bar{h}_{11}&=\bar{h}_++\bar{h}^* \\
\bar{h}_{22}&=-\bar{h}_++\bar{h}^* \\
\bar{h}_{33}&=\dfrac{2\omega^2}{c^2k^2-\omega^2}\,\bar{h}^*.
\end{split}
\end{equation}
 In order to analyze which kind of deformation is induced by each component on a sphere of test particles, we calculate the geodesic deviation equation, taken in the comoving frame
\begin{equation}\label{testparticles}
\dfrac{1}{c^2}\dfrac{d^2\xi^\mu}{dt^2}=R^\mu_{\; \nu \rho \sigma}  u^\nu u^\rho \xi^\sigma=R^\mu_{\; 00i}  \xi^i,
\end{equation}
where $\xi^\mu=\left(0,\xi^x,\xi^y,\xi^z \right)$ is a vector denoting the separation between two nearby geodesics and the Riemann tensor is constructed up to the first order in $\bar{h}_{\mu\nu}$, as given in \eqref{metricamezzo}. We fix the vector $\xi^\mu$ as
\begin{equation}
\xi^\mu=\left(0,\xi^x_{(0)}+\delta\xi^x,\xi^y_{(0)}+\delta\xi^y,\xi^z_{(0)}+\delta\xi^z\right),
\end{equation}
being $\xi^i_{(0)}$ the initial positions and $\delta\xi^i$ the displacements of order $O(h)$ induced by the gravitational wave. We compute \eqref{testparticles} up to order $h$ separately for each component of $\bar{h}_{\mu\nu}$, finding that:
\begin{enumerate}
	\item $\bar{h}_+$ is a standard plus mode in the $xy$ plane,
	\item $\bar{h}_{12}$ is a standard cross mode in the $xy$ plane.
	\item $\bar{h}_{13}$ is a cross mode in the $xz$ plane,
	\item $\bar{h}_{23}$ is a cross mode in the $yz$ plane,
	\item $\bar{h}^*$ is a superposition of two pure modes: a breathing mode\cite{flavio,g,h} in the $xy$ plane and a longitudinal mode along the $z$ axis.
	
\end{enumerate}
The ratio between the amplitudes of the anomalous polarizations and the standard ones (plus and cross polarization in the transverse plane) are found to be of order $k^2/m^2$.
\section{Models of macroscopic media}
Now we will give some quantitative estimates of $\epsilon_g$, without any intention of being too accurate, but merely realistic, in the characterization of the involved parameters. Let us begin with the case of a medium composed by molecules whose size is roughly the typical size of a binary system\cite{i,j,k,l,m}: we will set $L=30\, AU=4.4\cdot 10^{12}\,m$. In terms of the wavenumber, the condition $\lambdabar \gg L$ reads as $kL\ll 1$: we will fix $kL=0.05$, \textit{i.e.} $k=1.14 \cdot 10^{-14}\,m^{-1}$. The parameter $N$ changes considerably in different regions of the Galaxy, therefore we calculate $\epsilon_g$ for three different values of $N$: in $(i)$ we set $N=1\,pc^{-3}$ (our neighborhood), in $(ii)$ we choose the value $N=100\,pc^{-3}$ (inner region) and in $(iii)$ we fix $N=10^5\,pc^{-3}$ (Galaxy core).
We report the results obtained for $\epsilon_g$ and for the ratio $k^2/m^2$, that gives the entity of the deviation from $c$ in the group velocity and the ratio between the amplitudes of the anomalous polarizations and the standard ones:
\begin{center}
	\begin{itemize}
		\item [$(i)$]$\epsilon_g=1.89 \cdot 10^{40} \, kg \; m$   \qquad \qquad $k^2/m^2=2.27\cdot10^{-14}$\\
		\item [$(ii)$]$\epsilon_g=1.89 \cdot 10^{42} \, kg \; m$  \qquad \qquad $k^2/m^2=2.27\cdot10^{-12}$\\
		\item[$(iii)$]$\epsilon_g=1.89 \cdot 10^{45} \, kg \; m$  \qquad \qquad $k^2/m^2=2.27\cdot10^{-9}$.
	\end{itemize}
\end{center}
Now we consider the case of a material medium composed by molecules whose size is roughly comparable with the size of an open cluster\cite{n,o}: we set for $L$ the value $L=3\,pc=9.26\cdot 10^{16}\,m$. As before, we fix $kL=0.05$, \textit{i.e.} $k=5.40\cdot 10^{-19}\,m^{-1}$.
Our galaxy is estimated to contain about $100,000$ open clusters, mainly located in the central disc. We will calculate the density of molecules as the number of molecules ($10^5$) divided by the volume of the Galaxy, assuming it to be a disc with diameter of $45 \, kpc$ and thickness of the disc of about $0.6 \, kpc$ (roughly the size of the Milky Way \cite{p}). We obtain the value $N=3.55 \cdot 10^{-57} \, m^{-3}$.
We calculate $\epsilon_g$ and $k^2/m^2$:
\begin{itemize}
	\item []$\epsilon_g=8.13\cdot 10^{54}\, kg \; m \qquad \qquad k^2/m^2=2.21\cdot 10^{-8}$.
\end{itemize}
One can imagine a medium in which the density of clusters is $5$ or $50$ times greater than in our Galaxy: being the ratio $k^2/m^2$ linear in $N$, we expect for such a medium a magnitude of Macroscopic Gravity effects of order $10^{-7}\div 10^{-6}$. 
\section{Conclusions}
In this work we have demonstrated that the propagation of a gravitational wave in a molecular medium shows remarkable differences with respect to the vacuum case. By establishing a constitutive relation between the quadrupole tensor and the perturbative fields acting on the molecule, namely a vacuum gravitational wave $h_{\mu\nu}$ and the Newtonian potential of the molecule itself $\phi$, we have derived the law of motion for the wave inside the medium, a fourth order PDE: we have ruled out the possibility of damping and growing modes and determined that dispersion only occurs. We have outlined the fact that only the trace of the gravitational wave is still solution of d'Alembert equation, hence erasable with a gauge transformation. The remaining five degrees of freedom are not harmonic: they cannot be canceled out by a gauge transformation that preserves Hilbert gauge. This finding demonstrates that a gravitational wave in a matter medium possesses five degrees of freedom and it is able to induce longitudinal stresses in a sphere of test particles. We have showed that the quantity $k^2/m^2$ acts as a quantifier of two different observables: the deviation from $c$ in the group velocity and the ratio between the amplitudes of the anomalous polarizations and the standard ones. Lastly we have established some models of molecular medium, varying the values of the parameters $L$ and $N$: we have estimated the entity of Macroscopic Gravity effects, obtaining for $k^2/m^2$ values in the range $10^{-14} \div 10^{-6}$. We focused our analysis on two specific regions of wavelengths. In the first case (binaries) the signal is characterized by $\lambda \simeq 10^{13} \, m$, that is comparable with the scale of lengths to which the space interferometer LISA will be sensitive \cite{q,r}. In the second case (clusters) the wavelength of the signal is in the region $\lambda \simeq 10^{18} \, m$ and it is, in principle, detectable by experiments like IPTA \cite{s,t}. Despite the modest entity of the expected deviation from vacuum General Relativity, the integrated effect over very large distances can become significant and the propagation of gravitational waves can be considerably different from that of an electromagnetic counterpart, in a typical \textquotedblleft follow up\textquotedblright$\,$ scenario of multi-messenger astronomy.
\bibliographystyle{ws-procs975x65}


\end{document}